\documentstyle[fleqn,style2,12pt]{article}
\begin{document}
\renewcommand{\thepage}{}
\begin{titlepage}
\title{\Large The conical point in the ferroelectric six-vertex model}
\author{\Large
Dirk~Jan~Bukman and Joel~D.~Shore\\
{}~\\
\large
Department of Physics\\
Simon Fraser University\\
Burnaby, British Columbia\\
Canada V5A 1S6\\
{}~\\}
\maketitle
\date{}
\pagestyle{empty}
\begin{abstract}
\noindent We examine the last unexplored regime of the asymmetric
six-vertex model: the low-temperature
phase of the so-called ferroelectric model.
The original publication of the exact solution,
by Sutherland, Yang, and Yang, and various derivations and reviews
published afterwards, do not contain many details about this regime.
We study the exact
solution for this model, by numerical and analytical methods.  In
particular, we examine the behavior of the model in the vicinity of an
unusual coexistence point that we call the ``conical'' point.  This point
corresponds to additional  singularities in the free
energy that were not discussed in the original solution.
We show analytically that in this point many
polarizations coexist, and that unusual scaling
properties hold in its vicinity.
\end{abstract}
\end{titlepage}
\renewcommand{\thepage}{\arabic{page}}

\section{Introduction}

The six-vertex model is one of the few exactly solved models in
statistical mechanics.  Its origin lies in the problem of counting the
number of configurations of the hydrogen atoms in hydrogen-bonded
crystals, such as ice$^{(\cite{Pauling})}$ and
KH$_2$PO$_4$.$^{(\cite{Slater,LiebWu,Baxter})}$  In 1967, Lieb,
Yang, and Sutherland$^{(\cite{LYS})}$ obtained exact solutions for
various versions of the model, and the most general solution was
described by Sutherland, Yang, and Yang.$^{(\cite{SYY})}$
These publications contain virtually no
details, and later more explicit accounts were published by Lieb and
Wu,$^{(\cite{LiebWu})}$ and Nolden.$^{(\cite{Nolden,NoldenT})}$
The paper by Nolden contains a
detailed description of the exact solution, and explicit results for
some special cases where these can be obtained analytically.  However,
the low-temperature ferroelectric regime is not treated there.

A more recent application of the six-vertex model is as a model of
crystal surfaces.  Various restricted solid-on-solid models can be
mapped onto the six-vertex model, as was shown by van Beijeren for the
body-centered-cubic lattice,$^{(\cite{vBeijeren})}$ and by Jayaprakash and
Saam for the face-centered-cubic lattice.$^{(\cite{JayaprakashSaam})}$
This
last mapping in particular focuses attention on the low-temperature
ferroelectric regime of the six-vertex model.  In ref.\ \cite{JayaprakashSaam}
it was conjectured that there would be a discontinuity of slope at a
particular point in the crystal shape.  Since in this mapping the
crystal shape is, up to a scaling factor, given by
the shape of the free energy surface as a
function of applied fields,$^{(\cite{Andreev})}$ this implies the presence of
a singularity in the free energy at this point.  However, such a
singularity was not mentioned in the general solution of the six-vertex
model.$^{(\cite{SYY})}$  Our re-examination of the exact
solution in this regime, following the method presented in ref.\
\cite{Nolden}
shows that there is indeed a singularity at this point, and that it has
new and
unusual properties.  In this point, a one-parameter family of
polarizations coexists, and near it interesting scaling
properties are found.

The organization of the rest of the paper is as follows.  In Section
\ref{model}, we give a short, general discussion of the six-vertex model,
and briefly review the expressions from the exact solution that we use
further on.  Then, in Section \ref{deltalarger}, we apply those expressions
to a particular point in the phase diagram in the low-temperature
regime, where a completely analytical solution is possible.  The
behavior of the free energy and the order parameters at and around this
point is discussed.  In Section \ref{delta1} the same is done, in less
detail, at the transition between the low- and high-temperature regimes.
Conclusions and a summary of the results are given in Section
\ref{concl}.  The appendices contain calculational details that were
omitted from the main text of the paper.
A brief account of this work was published in ref.\ \cite{SB}.

\section{The ferroelectric six-vertex model}
\label{model}

A configuration of the six-vertex model is given by a covering of the
bonds of a square lattice with arrows, satisfying the ice rule:  Every
lattice point must have two incoming and two outgoing arrows.  There are
six possible vertices that satisfy this rule (see Fig.~\ref{vertices}).
To make
the model amenable to a solution by transfer matrix techniques
it is necessary to impose periodic boundary conditions in the horizontal
direction.  As a consequence, the number of down arrows is equal in each
row of bonds; this conservation law is essential for the exact solution of the
model.  The transfer matrix method used in the exact solution is
discussed in, {\it e.g.}, refs. \cite{LYS,LiebWu,Baxter}.

After assigning an energy to each of the vertices,
the partition function of the model
is simply given by the sum over all allowed configurations of the
Boltzmann factors of the configurations.  The usual way to write the
energies of the vertices is given in Fig.~\ref{vertices}.
The quantities $h$ and
$v$ are to be thought of as fields, acting on the horizontal and
vertical arrows respectively, favoring one orientation over the other.
(The name ``asymmetric six-vertex model'' refers to the fact that
these fields create an asymmetry between left and right and between up
and down arrows.)
The fact that vertices 5 and 6 have the same energy is no restriction:
since they are sinks and sources of horizontal arrows they have to occur
in pairs because of the periodic boundary conditions, so only the sum of
their energies is relevant.

The variables that are conjugate to the fields $h$ and $v$ are the
polarizations $x$ and $y$.  These are defined as $x=1-2f_{\leftarrow}$
and $y=1-2f_{\downarrow}$, where, {\it e.g.}, $f_{\leftarrow}$ is the fraction
of the
horizontal arrows pointing to the left.  $x$ and $y$ are order
parameters for the various ordered phases.  The free energy per vertex can be
viewed as either a function of the fields, $F(h,v)$, or of the
polarizations, $F(x,y)$.  The two are related by a two-dimensional Legendre
transformation
\begin{equation}
F(h,v)=\min_{x,y} \left\{F(x,y)-hx-vy\right\}
\label{Legendre}
\end{equation}
which implies $F(h,v)=F(x,y)-hx-vy$, with
the fields given by
\begin{equation}
h=\frac{\partial F(x,y)}{\partial x}, \mbox{~~~~} v= \frac{\partial
F(x,y)}{\partial y}
\label{derivs}
\end{equation}
or conversely
\begin{equation}
x=-\frac{\partial F(h,v)}{\partial h}, \mbox{~~~~} y= -\frac{\partial
F(h,v)}{\partial v}
\label{derivs2}
\end{equation}
The free energy that is calculated in the exact solution is actually
$F(h,y)$, because of the conservation law for the number of
down arrows.  This amounts to solving the model for  fixed values of
$h$ and $y$.  The relevant free energies $F(x,y)$ and $F(h,v)$
are then obtained from $F(h,y)$ through one-dimensional Legendre
transformations.

Most of the analysis in ref.\ \cite{Nolden} is valid for any values of the
vertex energies.  However, certain transformations that are applied do
depend on them; more precisely, they depend on the parameter
$\Delta  \equiv
\frac{1}{2}(\eta+\eta^{-1}- e^{2\beta\varepsilon})$, where $\eta\equiv
e^{\beta\delta}$, and $\beta = 1/k_B T$ is the inverse temperature.
The discussion
in ref.\ \cite{Nolden} is restricted to the case $\Delta < 1$;
we will focus on $\Delta
\geq 1$.  There is a different transformation for $\Delta=1$ and for
$\Delta>1$; both are given in ref.\ \cite{SYY}.  As a consequence, various
quantities and
functions that occur in the exact solution have different forms in these
two cases; they are given in table~\ref{ftable}
(the expressions for the other ranges of $\Delta$ are given in ref.\
\cite{LiebWu,Nolden}).
$\Delta > 1$ corresponds to the low-temperature phase of models that are
``ferroelectric'', {\it i.e.} that have $\varepsilon < \delta/2$, so that
the vertices 1 and 2, with a non-zero net
polarization, have lower energies (for $h=v=0$)
than the vertices 5 and 6 with zero net polarization
(we choose $\delta\geq 0$ without loss of generality).
By varying $T$ at fixed $\delta $ and $\varepsilon$,
a transition to a high-temperature phase (with $\Delta <1$) takes
place at a critical temperature $T=T_c$ (corresponding to $\Delta =1$).

We now give a brief review of the main results in Nolden's
paper$^{(\cite{Nolden})}$ that we will use; for derivations and more details
the reader should
consult that paper.
For fixed $\varepsilon$, $\delta$, and $T$, the exact solution
provides expressions for the free energy, and various other quantities
of interest, like the fields and the polarizations, as a function of two
parameters, $a$ and $b$.  Different
values of $a$ and $b$ then correspond to different points in the phase
diagram, {\it i.e.}\ different values of $h$, $v$, $x$, and $y$.
The first step is to find the function $R(u)$ from the
integral equation
\begin{equation}
R(u)+\frac{1}{2\pi}\int_{-a}^{a} K(u-v) R(v)dv=\xi(u)
\label{Req}
\end{equation}
$R$ is a function of $u$, a real parameter in the interval $[-a,a]$;
it also depends on the values of $a$ and $b$.\footnote{Here and in what
follows, explicit dependence
of functions on $a$ and $b$ has been omitted for brevity.}
This equation results
from the consistency relations for the wavenumbers in the Bethe ansatz.
Once $R$ is known, the free energy per site can be obtained as the largest
eigenvalue of the transfer matrix,
\begin{equation}
-\beta F(h,y)=\max_{\rm R,L} \left[\pm\frac{1}{2}(\ln\eta+2\beta h)
+\frac{1}{2\pi}\int_{-a}^{a} \Phi^{\rm R,L}(u)R(u)du\right]
\label{feeq}
\end{equation}
with upper or lower sign for the right (R) or left (L) eigenvalue,
respectively.
Expressions for $h$ and $y$ can be obtained from the ``generalized
normalization''
\begin{equation}
\frac{\pi}{2}(1-y) -2i \beta h = p^0(a)-\frac{1}{2\pi}\int_{-a}^{a}
\Theta(a-v) R(v)dv
\label{yheq}
\end{equation}
To find $v$, $x$, and the relevant free energies,
$F(h,y)$ must be differentiated with respect to $h$ and $y$, on which
variables it depends only implicitly through $a$ and $b$.  All this is
described in ref.\ \cite{Nolden}, and that treatment applies to the
ferroelectric case without any modification.

For general values of $a$ and $b$,
the only way to solve these equations is numerically; the integral
equation can be solved with standard techniques (See
ref.\ \cite{NumericalRecipes}, Chap.\ 18),
and all quantities can be calculated.
The ranges of $a$ and $b$ in table~\ref{ftable} lead to that part of the
phase diagram where $y\geq 0$.  The other half can be obtained through a
simple symmetry operation.  An analytic solution is possible
for $a=0$ and $a=-\pi$, and an expansion around these points can also be
made.  The case $a=0$ was examined in ref.\ \cite{JayaprakashSaam};
it corresponds to second order boundaries between completely and
incompletely polarized phases in the phase diagram.
We will focus on the case $a=-\pi$ , which corresponds to
the two ``conical'' points.

\section{Analytic solution for $\Delta > 1$}
\label{deltalarger}
In the case that $a=-\pi$, the integral equations in the previous
section can be solved analytically by means of Fourier series, as was
done for $a=\pi$ in the case $\Delta < -1$ in ref.\ \cite{Nolden}.  It is
then also possible to calculate an explicit expression for the free energy,
and to examine its behavior around the point $a=-\pi$.

\subsection{Finding $R(u)$}

The first step in the solution is to solve Eq. (\ref{Req}).  To find
the behavior both at and around $a=-\pi$, we write the
solution $R$ as an expansion in\footnote{This
$\epsilon$ should not be confused with the energy
$\varepsilon$ introduced earlier.}
$\epsilon=\pi+a$:
\begin{equation}
R(u)=R_0(u)+\sum_{m=1}^{\infty} \epsilon^m \delta R_m(u)
\label{Rexp}
\end{equation}
All integrals of the form $\int_a^{-a} f(v) dv$ can be expanded in
$\epsilon$ as follows:
\begin{eqnarray}
\int_{-\pi+\epsilon}^{\pi-\epsilon} f(v)dv &=&\int_{-\pi}^{\pi}
f(v)dv \nonumber\\
&&+\sum_{m=1}^{\infty} \frac{1}{m!}\left\{(-\epsilon)^m
f^{(m-1)}(\pi) - \epsilon^mf^{(m-1)}(-\pi)\right\}
\label{intexp}
\end{eqnarray}
Using this in Eq. (\ref{Req}), and substituting the expansion
(\ref{Rexp}) everywhere, we get a set of integral equations,
one for each order in $\epsilon$, containing $R_0$ and the $\delta
R_m$. All functions $R_0$ and $\delta R_m$ are defined on the interval
$u\in [-\pi,\pi]$, and all integrals are over the same interval.
Since $K$ and $\xi$ have period $2\pi$, the solutions $R_0$ and
$\delta R_m$ also have this periodicity.  (Note that by this periodicity
terms with even $m$ in the expansion (\ref{intexp}) vanish.  The terms
with odd $m$ reduce to twice the value at $v=\pi$.)  This
periodicity makes it possible to solve for the functions $R_0$ and
$\delta R_m$ by applying a Fourier expansion to the equation. We will
now examine the first five orders in $\epsilon$, and solve
for $R_0$ and the first four $\delta R_m$.

The Fourier components of, {\it e.g.}, $R_0(u)$  are defined by
\begin{equation}
R_0(u)=\sum_{n=-\infty}^{\infty} \widehat{R}_ne^{-inu}
\label{fc}
\end{equation}
and similarly for $\delta R_m$, $K$ and $\xi$.
The Fourier coefficients $\widehat{K}_n$ are
the same as those given in ref.\ \cite{Nolden}, and
$\widehat{\xi}_n $ is calculated in~\ref{xi}:
\begin{eqnarray}
\widehat{K}_n&=& e^{-2\nu|n|}\nonumber\\
\widehat{\xi}_n&=&\left\{
\begin{array}{ll}
\left\{ \begin{array}{ll}
0 & (n \geq 0)\\
2e^{bn}\sinh(n\nu) & (n<0)
\end{array} \right. & (b > \nu)\\
\left\{ \begin{array}{ll}
-2e^{bn}\sinh(n\nu) & (n \geq 0)\\
0 & (n<0)
\end{array} \right. & (b< -\nu)
\end{array} \right.
\end{eqnarray}

Eq. (\ref{Req}) to order $\epsilon^0$ is
\begin{equation}
R_0(u)-\frac{1}{2\pi}\int_{-\pi}^{\pi} K(u-v)R_0(v,b) dv =\xi(u)
\label{ozero}
\end{equation}
Fourier transforming this gives
\begin{equation}
\widehat{R}_n[1-\widehat{K}_n]=\widehat{\xi}_n
\label{ozerohat}
\end{equation}
so, for $n\neq 0$, $\widehat{R}_n$ is given by
\begin{eqnarray}
\widehat{R}_n&=&\left\{
\begin{array}{ll}
\left\{ \begin{array}{ll}
0 & (n > 0)\\
-e^{(b-\nu)n}& (n<0)
\end{array} \right. & (b > \nu)\\
\left\{ \begin{array}{ll}
-e^{(b+\nu)n}& (n > 0)\\
0 & (n<0)
\end{array} \right. & (b< -\nu)
\end{array} \right.
\label{fred}
\end{eqnarray}
For $n=0$, (\ref{ozerohat}) reduces to $0=0$, so $\widehat{R}_0$ is not yet
determined.  The Fourier series (\ref{fc}) for $R_0$ can be summed, and we find
\begin{equation}
R_0(u)
=\widehat{R}_0 - \frac{e^{\nu\mp b\pm iu}}{1-e^{\nu\mp b\pm iu}}
\label{rzero}
\end{equation}
with the upper (lower) signs for $b>\nu$ ($b<-\nu$).
$\widehat{R}_0$ will be determined by the next order in $\epsilon$.

Next, for $\epsilon^1$ we find
\begin{equation}
\delta R_1(u) -\frac{1}{2\pi}\int_{-\pi}^{\pi} K(u-v)\delta R_1(v)dv=
-\frac{1}{\pi}\left[K(u-\pi)R_0(\pi)\right]
\end{equation}
After Fourier transforming,
and using $K(u-\pi)=\sum_n(-1)^n \widehat{K}_n e^{-inu}$, we have
\begin{equation}
(\widehat{\delta R_1})_n [1-\widehat{K}_n] = -\frac{(-1)^n}{\pi}
\widehat{K}_n R_0(\pi)
\end{equation}
Since $\widehat{K}_0=1$, there is a problem unless the r.h.s. is
zero for $n=0$.  Thus $R_0(\pi)$ must be zero.  This requirement
fixes $\widehat{R}_0$, giving
\begin{equation}
\widehat{R}_0=\left\{
\begin{array}{ll}
-\frac{e^{\nu-b}}{1+e^{\nu-b}} & (b > \nu)\\
-\frac{e^{\nu+b}}{1+e^{\nu+b}} & (b<-\nu)
\end{array}\right.
\label{rzerohat}
\end{equation}
Once we put $R_0(\pi)=0$ it is obvious that
$(\widehat{\delta R_1})_n=0$ for $n\neq 0$.
$(\widehat{\delta R_1})_0$ will be determined in the next order.

We continue in this spirit, finding expressions for the next
terms:
\begin{eqnarray}
(\widehat{\delta R_1})_0 &=&0\nonumber\\
\delta R_2(u)&\equiv&\delta R_2 = -\frac{1}{6}R_0''(\pi)\nonumber\\
(\widehat{\delta R_3})_n &=&
\left\{
\begin{array}{ll}
-\frac{in(-1)^n \widehat{K}_n R_0'(\pi)}{
3\pi(1-\widehat{K}_n)}&(n\neq 0)\\
0 &(n=0)\\
\end{array}\right.
\nonumber\\
\delta R_4(u)&\equiv&\delta R_4=-\frac{1}{120}R_0^{(4)}(\pi)
\end{eqnarray}

The above expressions contain the first, second and fourth derivatives
of $R_0(v)$ evaluated at $v=\pi$.  These are obtained from Eq.
(\ref{rzero}),
\begin{eqnarray}
R_0'(\pi)&=& \pm i\frac{e^{\nu\mp b}}{(1+e^{\nu\mp b})^2}\nonumber\\
R_0''(\pi)&=& -\frac{e^{\nu\mp b}(1-e^{\nu\mp b})}{(1+e^{\nu\mp b})^3}
\nonumber\\
R_0^{(4)}(\pi)&=& \frac{e^{\nu\mp b}(1-11e^{\nu\mp b}+11e^{2(\nu\mp b)}-
e^{3(\nu\mp b)})}{(1+e^{\nu\mp b})^5}
\end{eqnarray}
with the upper (lower) signs for $b>\nu$ ($b<-\nu$).

To recapitulate, the total expression for $R(u)$ is
\begin{equation}
R(u)=R_0(u)+\epsilon^2\delta R_2 +\epsilon^3 \delta R_3(u)
+\epsilon^4 \delta R_4+{\cal O}(\epsilon^5)
\end{equation}
where $\delta R_2$ and $\delta R_4$ are
real and independent of $u$, and $\delta R_3(u)$ is imaginary.

\subsection{Calculating $y$ and $h$}

Having determined $R$, we can now determine $y$ and $h$ using
(\ref{yheq}).  We will first
calculate the values $y_0$ and $h_0$, the first terms in the expansion
in $\epsilon$, and then find terms up to order $\epsilon^4$ from a
slightly simplified equation.
{}From table~\ref{ftable}, $p^0(-\pi)$
is given by
\begin{equation}
p^0(-\pi)=-i\ln\left[\frac{e^{\nu}+e^b}{e^{\nu+b}+1}\right]
\end{equation}
The second term on the right hand side of (\ref{yheq}) is
\begin{equation}
\frac{1}{2\pi}\int_{-\pi}^{\pi}\Theta(-\pi-v) R_0(v) dv =
\frac{1}{2\pi}\sum_n \widehat{R}_n \int_{-\pi}^{\pi} \Theta(-\pi-v )
e^{-inv} dv
\end{equation}
The integrals $I_n\equiv-\int_{-\pi}^{\pi} \Theta(-\pi-v )
e^{-inv} dv$ are calculated in~\ref{eye}.  They turn out to be
\begin{equation}
I_n=\left\{
\begin{array}{ll}
(-1)^n \frac{2\pi i}{n}(1-e^{-2n\nu}) & (n>0)\\
2\pi^2 & (n=0)\\
(-1)^n \frac{2\pi i}{n}(1-e^{2n\nu}) & (n<0)
\end{array}\right.
\end{equation}
For the aforementioned term we now have
\begin{equation}
-\frac{1}{2\pi}\sum_n I_n \widehat{R}_n
= -\pi \widehat{R}_0 \pm i[\ln(1+e^{\nu\mp b}) - \ln(1+e^{-\nu\mp
b})]
\end{equation}
with the upper (lower) signs for $b>\nu$ ($b<-\nu$).
The final result then is
\begin{equation}
\frac{\pi}{2}(1-y_0)
-2i\beta h_0
=\left\{
\begin{array}{ll}
-\pi\widehat{R}_0 + i\nu & (b>\nu)\\
-\pi\widehat{R}_0 -i\nu  & (b<-\nu)
\end{array}\right.
\end{equation}
So we find, using Eq.\ (\ref{rzerohat}),
\begin{eqnarray}
y_0 &=& \left\{
\begin{array}{ll}
\tanh(\frac{b-\nu}{2}) & (b>\nu)\\
-\tanh(\frac{b+\nu}{2}) & (b<-\nu)
\end{array}\right.
\nonumber\\
h_0 &=& \left\{
\begin{array}{ll}
-\nu/2\beta & (b>\nu)\\
\nu/2\beta & (b<-\nu)
\end{array}\right.
\end{eqnarray}

Now we use Eq.\ (2.28) from ref.\ \cite{Nolden} to find
the next orders in $\epsilon$.
This equation is
\begin{eqnarray}
-\frac{\pi}{2}\partial_a y -2i\beta\partial_a h &=&R(a)-\frac
{1}{2\pi}\Theta(2a)R(-a) \nonumber\\
&& + \frac{1}{2\pi}\int_{a}^{-a}\Theta(a-v)\partial_a R(v)dv
\end{eqnarray}
Substituting the expansions for $R$, $\partial_a R$,
expanding all occurrences of $a$ to order $\epsilon^3$, and using
\ref{eye}, we find
\begin{equation}
-\frac{\pi}{2}\partial_a y -2i\beta\partial_a h =
\epsilon\frac{\pi}{3}R_0''(\pi) +\epsilon^2\frac{1}{\pi}
R_0'(\pi)+\frac{\pi}{30}\epsilon^3 R_0^{(4)}(\pi)
\end{equation}
Since $R_0''$ and $R_0^{(4)}$ are
real and $R_0'$ imaginary, we have
\begin{eqnarray}
\partial_a y &=&-\frac{2}{3} \epsilon R_0''(\pi)-\frac{1}{15}
\epsilon^3 R_0^{(4)}(\pi)+{\cal O}(\epsilon^4)
\nonumber\\
2\beta \partial_a  h&=& i\epsilon^2\frac{R_0'(\pi)}{\pi}
+{\cal O}(\epsilon^4)
\end{eqnarray}

Putting all this together, we now have power series in $\epsilon$ for $y$
and $h$:
\begin{eqnarray}
y&=&y_0+\epsilon^2 y_2 +\epsilon^4 y_4 +{\cal O}(\epsilon^5)
\nonumber\\
h&=&h_0+\epsilon^3 h_3 +{\cal O}(\epsilon^5)
\label{coefexp}
\end{eqnarray}
with
\begin{eqnarray}
y_0&=&\pm\tanh\left(\frac{b\mp \nu}{2}\right)
\nonumber\\
y_2&=& \frac{1}{3}\frac{e^{\nu\mp b}(1-e^{\nu\mp b})}{(1+
e^{\nu\mp b})^3}\nonumber\\
y_4&=& -\frac{1}{60} \frac{e^{\nu\mp b}(1-11e^{\nu\mp b}+11e^{2(\nu\mp b)}-
e^{3(\nu\mp b)})}{(1+e^{\nu\mp b})^5}
\nonumber\\
h_0&=&\mp\frac{\nu}{2\beta}
\nonumber\\
h_3&=&
\mp\frac{1}{6\pi\beta}\frac{e^{\nu\mp b}}{(1+e^{\nu\mp b})^2}
\label{coef1}
\end{eqnarray}
with the upper (lower) signs for $b>\nu$ ($b<-\nu$).

\subsection{Calculating $x$ and $v$}
\label{symms}

To rigorously calculate expressions for $x$ and $v$, and the Legendre
transforms $F(x,y)$ and $F(h,v)$, one has to construct the derivatives
of $F(h,y)$ by applying the chain rule.  This is what is done in
ref.\ \cite{Nolden}.  However, by making a few assumptions, it is possible to
derive the same expressions from the symmetry properties of the exact
solution. The assumptions can be checked against known exact results, in
which case they turn out to hold.

The first step is to realize that, instead of calculating the free energy
as a function of $h$ and $y$, one might just as well calculate it as a
function of another pair of variables, for example $v$ and $x$.  The
results will be exactly the same as before, except for the change of
variables, and the region of the phase diagram that one finds the
solution for.  The preceding calculation gives $h$, $y$, and $F(h,y)$
expressed in terms of the parameters $a$ and $b$.  The solution holds
for $y\geq 0$, which corresponds to regions II and III (for
$b\geq\nu$), and region IV (for $b\leq-\nu$) in
Fig.~\ref{pdhv}.\footnote{In this section we are only concerned with the
so-called incompletely polarized part of the phase diagram, which
consists of the regions I -- VI.  This is because this is the only part
that is mapped out by varying $a$ and $b$.  In the rest of the phase
diagram, $x$ and $y$ are frozen at their extreme values $\pm 1$. This is
discussed in detail in section~\ref{disc}.}
The solution for $y$ is of the form
\begin{equation}
y=\left\{
\begin{array}{ll}
Y_1(a,b) &\mbox{~~~~in regions II and III}\\
Y_2(a,b)& \mbox{~~~~in region IV}\\
\end{array}\right.
\end{equation}
where $Y_1$ and $Y_2$ are given functions.  If, on the other hand, one
calculates $F(v,x)$, one finds exactly the same solution, but now $v$
and $x$ play the roles of $h$ and $y$, and the parameters are now
called
$\tilde{a}$ and $\tilde{b}$ to distinguish them from the other set of
parameters.  The solution is now for that part of the phase diagram
where $x\geq 0$, regions V and IV (for $\tilde{b}\geq\nu$), and region
III (for $\tilde{b}\leq-\nu$).  The expression for $x$ is just the same
as the one found previously for $y$,
\begin{equation}
x=\left\{
\begin{array}{ll}
Y_1(\tilde{a},\tilde{b}) &\mbox{~~~~in regions V and IV}\\
Y_2(\tilde{a},\tilde{b}) &\mbox{~~~~in region III}\\
\end{array}\right.
\end{equation}
with the same functions $Y_1$ and $Y_2$.  We now have expressions for
both $y$ and $x$ in regions III and IV, only they are given in terms  of
two different sets of parameters, $a,b$ and $\tilde{a},\tilde{b}$.
We need a relation between them to find $x$ and $y$ expressed in $a$ and
$b$.  A simple ansatz is to assume that $a=\tilde{a}$, and that
$b$ and $\tilde{b}$ are related linearly.  Since the line $b=\phi_0$
corresponds to $\tilde{b}=-\nu$, and $b=\infty$ to $\tilde{b}=-\infty$,
this would give $\tilde{b}=-b +\phi_0-\nu$.  Substituting this,
we find for $x$
\begin{equation}
x=\left\{
\begin{array}{ll}
X_1(a,b)=Y_2(a,-b+\phi_0-\nu) &\mbox{~~~~in region III}\\
X_2(a,b)=Y_1(a,-b+\phi_0-\nu) &\mbox{~~~~in region IV}\\
\end{array}\right.
\end{equation}
If we further assume that the expression for $x$ in region III also
holds in region II, as is true for $y$, we now have expressions for $x$
throughout the region $y\geq 0$.  The expression for $x$ for $b\geq \nu$
can simply be obtained by taking $y$ for $b\leq-\nu$ and substituting
$-b+\phi_0-\nu$ for $b$.  Similarly, $x$ for $b\leq-\nu$ follows from
the expression for $y$ for $b\geq \nu$ with the same substitution.
The same prescription can be used to obtain $v$ from $h$.

It is possible, to a certain extent, to check the assumptions we made.
As a first check, we have calculated $x$ and $v$ to zeroth order in
$\epsilon$ the rigorous way, obtaining the same results as above.
Second, from the symmetry properties we can also derive consistency
relations for the expressions for $y$ in various regions.  For example,
we can repeat the above reasoning for $F(-v,-x)$.  This gives the
region of the phase diagram where $x\leq0$, regions I and II (for
$\bar{b}\geq \nu$), and VI (for $\bar{b}\leq-\nu$).  The expressions for
$x$ are now
\begin{equation}
x=\left\{
\begin{array}{ll}
-Y_1(\bar{a},\bar{b}) &\mbox{~~~~in regions I and II}\\
-Y_2(\bar{a},\bar{b}) &\mbox{~~~~in region VI}\\
\end{array}\right.
\end{equation}
If we make the same assumptions as before, $\bar{a}=a$ and
$\bar{b}=\nu+\phi_0-b$.  Then
in region II, $x$ is given by $-Y_1(a,-b+\nu+\phi_0)$.  But in that
region $x$ is also given by $Y_2(a,-b+\phi_0-\nu)$.  So to be consistent
the expressions for $y$ have to satisfy $Y_1(a,-b+\nu+\phi_0)=-
Y_2(a,-b+\phi_0-\nu)$.  Checking the expressions for the first few orders in
$\epsilon$ (\ref{coef1}) shows that this is indeed true,
giving confidence
that the assumptions we made are justified.  In addition, the numerical
results confirm the expressions for $x$ and $v$ found in this way.

\subsection{Finding the free energy}

Now we are in a position to calculate $F(h,y)$ from (\ref{feeq}).
We need to know the integrals
\begin{equation}
\int_{-\pi}^{\pi} \Phi^{\rm R,L} (u) e^{-inu}
du
\end{equation}
These can be expressed in terms of integrals of the form
\begin{equation}
J_n = \int_{\bigcirc} dz \left[ \ln\left(\frac{z-A}{z-B}\right) + C
\right]z^{n-1}
\end{equation}
for $n\geq 0$.  These are calculated in~\ref{jay}.

The details of the calculation of the zeroth order part $F_0$ of the
free energy as a function of $y$ and $h$
can be found in~\ref{fe}.  There it is shown that the right
eigenvalue is largest for $b<-\nu$ and $b> \phi_0$, and the left
eigenvalue is largest for $\nu<b<\phi_0$.  Even though at $b=\phi_0$ the
two eigenvalues cross, the analytic form of the
expression for the free energy $F_0$ is the same on either side of
$b=\phi_0$.
(In fact, this remains true at least up to order $\epsilon^3$).
In~\ref{fe} we find
\begin{equation}
-\beta F_0(h,y)= \frac{\beta\delta}{2} \mp\nu(\widehat{R}_0+\frac{1}{2})
=\frac{\beta\delta}{2} \mp\frac{\nu}{2}y_0
\end{equation}
with the upper (lower) signs for $b>\nu$ ($b<-\nu$).
Note that for $b>\nu$, $v_0=\partial F/\partial
y=\nu/2\beta$, so $a=-\pi$ and $b>\nu$ corresponds to
the point $(h_0,v_0)=
(-\nu/2\beta, \nu/2\beta)$.  Similarly, $a=-\pi$ and $b<-\nu$ corresponds
to $(h_0,v_0)=(\nu/2\beta, -\nu/2\beta)$.

To find the next two orders of $F(h,y)$ in $\epsilon$ we expand Eq.
(\ref{feeq}) as usual, to order $\epsilon^3$, which gives
\begin{equation}
-\beta(F-F_0)=
\pm\frac{1}{6} \epsilon^2\nu R_0''(\pi) \pm
\frac{\epsilon^3}{6\pi}\frac{e^{\nu\mp b}}{(1+e^{\nu\mp b})^2}
\frac{1-e^{\pm\sigma-\nu}}{1+e^{\pm\sigma-\nu}}
\end{equation}
with the upper (lower) signs for $b>\nu$ ($b<-\nu$),
and $\sigma \equiv b-\phi_0+\nu$.
The general structure of $F(h,y)$ is thus found to be
\begin{equation}
F(h,y)=F_0+\epsilon^2 F_2+\epsilon^3 F_3 +{\cal O}(\epsilon^4)
\end{equation}
The coefficients $F_n$ can be expressed in terms of the coefficients in
the expansions of $h,v,x$, and $y$ (see Eq. (\ref{coef1})):
\begin{eqnarray}
F_0&=&-\frac{\delta}{2}+v_0y_0\nonumber\\
F_2&=&v_0y_2\nonumber\\
F_3&=&-h_3x_0
\end{eqnarray}
Since the Legendre transforms of the free energy are given by $F(x,y)=
F(h,y)+xh$ and $F(h,v)=F(h,y)-vy$, we also have
\begin{eqnarray}
F(x,y)&=& -\frac{\delta}{2} +v_0y_0+h_0x_0 +(v_0y_2+h_0x_2)\epsilon^2
+{\cal O}(\epsilon^4)\label{fexyeq}\\
F(h,v)&=& -\frac{\delta}{2} -(h_3x_0+v_3y_0)\epsilon^3 +
{\cal O}(\epsilon^4)\label{fehveq}
\end{eqnarray}

\subsection{Discussion}
\label{disc}

The overall phase diagram as a function of $h$ and $v$ can be obtained
numerically; its central region is shown as the base plane in
Fig.~\ref{fehv}.  It contains four phases
where the polarizations $(x,y)$ are ``frozen in'' at their extremal
values $(\pm 1, \pm 1)$.  Two of these phases,
$(1,1)$ and $(-1,-1)$, meet along the line $h=-v$.  $x$ and $y$ are
discontinuous across this line, so it represents a first order phase
transition.  In between the four frozen regions are two
incompletely polarized phases, where $x$ and $y$ change
continuously with $h$ and $v$.  They are separated from the frozen
phases by continuous transitions, which have been shown to be
of the Pokrovsky-Talapov (PT) type.$^{(\cite{JayaprakashSaam,PT})}$
The points where two PT lines meet
the first order line are the ``conical'' points.

To show the relation between the values of the parameters $a$ and $b$,
and the phase diagram in terms of $h$ and $v$,
several lines of constant $a$ and $b$ have been drawn in
Fig.~\ref{lines}. This relation is shown only for the
part of the phase diagram with $b \ge (\nu+\phi_0)/2$.
This gives one quarter of the phase
diagram; the rest of the diagram is then obtained by symmetry.
We restricted our computation to this region to avoid certain
numerical difficulties that arise near $b=\pm\nu$:  At
$b=\pm\nu$, $R(u)$ has a pole and this makes the solution of the integral
equations increasingly difficult as $b\to\pm\nu$.
An additional difficulty is that at $b=\phi_0$,
$\Phi^{R,L}(u)$ have poles.  This does not affect the integral
equations themselves, but does make the integrals involved in the
calculation of $F(h,y)$, $x$, and $v$ increasingly difficult as $b\to\phi_0$.
We have dealt with this problem by using an integration routine with
adaptive stepsize control (See ref.\ \cite{NumericalRecipes}, Chap.\ 4), as
opposed to the simple Gaussian quadrature used for solving the integral
equations and determining the less--sensitive integrals.  However,
near $b=\phi_0$ the integration does become quite slow.

Fig.~\ref{fehv} also shows the free energy surface $F(h,v)$ superimposed on
the phase diagram.  $F(h,v)$ is purely linear in $h$ and $v$ in the
frozen phases, reflecting the fact that $(x,y)=-(\partial F/\partial h,
\partial F/\partial v)$ is fixed.  In the incompletely polarized phases,
$x$ and $y$ change continuously, so $F(h,v)$ is smoothly curved.  Along
the PT lines the curved part of $F$ joins smoothly onto the linear part,
since the singularity in $F(h,v)$ corresponding to the (second order) PT
transition leaves the first derivative continuous.
Along the first order
transition line there {\em is} a jump in slope, so that a ridge is formed.  At
the two conical points the slope of $F$ depends on the way in which the
point is approached, as will be shown later on.

We now turn to the phase diagram in $x$-$y$ space, Fig.~\ref{fexy}.
Here the four frozen phases correspond to the corner points $
(x,y)=(\pm 1, \pm
1)$.  The incompletely polarized phases take up the rest of the phase
diagram outside the lines $\ell_+$ and $\ell_-$.
The two conical points now correspond
to the two lines $\ell_+$ and $\ell_-$, which are given by $(x_0(b),y_0(b))$
for
$b>\nu$ and $b<-\nu$, respectively (see Eq. (\ref{coef1})).
By eliminating $b$ in Eq. (\ref{coef1})
we find that the lines $\ell_{\pm}$ are given by
\begin{equation}
y_0 = \frac{x_0\pm\tanh[(\phi_0-\nu)/2]}{1\pm x_0\tanh[(\phi_0-\nu)/2]}
\end{equation}
{}From
Eq. (\ref{fexyeq}) we see that on, {\it e.g.}, $\ell_+$, where $a=-\pi$ and
$b>\nu$, the free energy is $F(x,y)=-\delta/2 +v_0y_0+h_0x_0 = -\delta/2
+(\nu/2\beta) (y_0-x_0)$.  This means that the slope of $F(x,y)$, which is
equal to $(h,v)= (-\nu /2\beta,\nu /2\beta)$, is the same everywhere along
$\ell_+$, and that every $(x_0,y_0)$ on $\ell_+$ corresponds to the same
point in $(h,v)$, the conical point.  So, all the polarizations $(x_0,y_0)$
are stable in the conical point.  This is similar to the coexistence of
a one-parameter family of magnetizations $\vec{m}$ with $|\vec{m}| =
m_0(T)$ in a zero-field XY model for $T< T_c$.  However, there the
coexistence is due to an obvious symmetry in the Hamiltonian, while in
the six-vertex model the symmetry that allows the polarizations to
coexist is not an obvious symmetry in the Hamiltonian and therefore
appears to be generated spontaneously.

It can be shown that different values
of $b$ correspond to entering the conical point with different angles from
the incompletely polarized region; if $b$ is kept fixed as $a\rightarrow
-\pi$, the conical point is approached from an angle
\begin{equation}
\theta=\arctan\left[\frac{\cosh(b-\nu)-\cosh(b-\phi_0)}{2+\cosh(b-\nu)+\cosh(b
-\phi_0)}\right]
\label{angle}
\end{equation}
with respect to the line $h=-v$.  Since every
different $b$ leads to a different value of $(x,y)$, and $(x,y)$ gives
the slope of $F(h,v)$, the slope in the conical point depends on the
angle at which it is entered. This gives $F(h,v)$ a geometry
similar to the tip of a cone at this point, hence the name ``conical''
point.

The exact solution does not give any states with polarizations in the
regions $C^{\pm}$ between $\ell_+$ and $\ell_-$.  Thus, no pure
equilibrium states with those polarizations exist, and strictly speaking
the free energy $F(x,y)$ is not defined in this region.  However, a
state with an {\em average} polarization $(x,y)$ in, {\it e.g.}, $C^+$ can be
formed as a mixture of states on $\ell_+$, properly weighted to give the
right average polarization.  This leads to a free energy that is also an
average of the free energies of the states in the mixture, which means
that it is linear in $x$ and $y$.  In this way, the regions $C^{\pm}$
are interpreted as describing mixtures of the coexisting phases on
$\ell_{\pm}$, similar to states that are a mixture of coexisting gas and
liquid phases at a gas-liquid phase transition.  The slope of the free
energy surface $F(x,y)$
is then given by $(h,v)=(-\nu/2\beta,\nu/2\beta)$ in
$C^{+}$, and $(h,v)=(\nu/2\beta,-\nu/2\beta)$ in $C^{-}$.  This matches
the slopes on the lines $\ell_+$ and $\ell_-$,
so that the first derivative of
the free energy $F(x,y)$ is continuous across these lines.
Its higher derivatives are not continuous, however, and
the free energy has
singularities at the lines $\ell_{\pm}$,
since an analytic function
cannot be matched to a purely linear function in a regular way.
This form of the free energy also means that
the fields corresponding to the regions $C^{\pm}$ are the same as those
for the lines $\ell_{\pm}$, {\it i.e.} the mixtures described by $C^{\pm}$
occur in the conical points.  The two linear parts of $F(x,y)$ meet
along the line $x=y$, forming a groove.  This groove corresponds to the
coexistence of the two frozen phases along the first order transition
in Fig.~\ref{fehv},
and it was already mentioned in ref.\ \cite{SYY}.  However, it is stated there
that the groove is the only set of points in $x,y \in (-1,1)$ where
$F(x,y)$ is non-analytic, while in fact there are additional
non-analyticities along the lines $\ell_{\pm}$.

To examine the nature of the singularities in the free energy we now
specialize to the case $b=(\nu+\phi_0)/2$, so that we are on the lines
$y=-x$ and $v=-h$, moving from the incompletely polarized phase towards
the conical point as $\epsilon \rightarrow 0$.
The behavior along this line should be qualitatively the same as along
any other line emanating from the conical point;
however, it is much easier to deal with
just two variables instead of four.  Solving for $\epsilon$ in $\delta
v\equiv v-v_0=\epsilon^3 v_3 +{\cal O}(\epsilon^5)$, and
substituting this in the expression for $y$ (Eq. \ref{coef1}), we find that
\begin{equation}
y(v)= y_0 +\frac{y_2}{v_3^{2/3}} \delta v^{2/3} +{\cal O}(\delta
v^{4/3})
\label{joel1}
\end{equation}
So, as $\delta v$ goes to zero, $y$ approaches its value of $y_0$ in the
conical point with a slope that diverges like $\delta v^{-1/3}$.
This value is in agreement with recent
work$^{(\cite{DenNijs,Private})}$ which shows that certain
(1+1)--dimensional surface growth models map onto the conical point in the
$(h,v)$
phase diagram, and that
Kardar-Parisi-Zhang (KPZ)$^{(\cite{KPZ})}$ scaling holds.  In particular, the
exponent of $2/3$ in Eq.\ (\ref{joel1}) is given
by $1/z$,$^{(\cite{Private})}$ where $z=3/2$ is the
dynamic exponent for the KPZ universality class.
Relation (\ref{joel1})
can be integrated to give $F(v)$, since $\partial F(v)/\partial v=
-2y$ (the factor 2 comes from the fact that $x=-y $ and $h=-v$).  This
gives
\begin{equation}
F(v) = -\frac{\delta}{2}-2y_0\delta v -\frac{6y_2}{5v_3^{2/3}}\delta
v^{5/3} +{\cal O}(\delta v^{7/3})
\label{joel2}
\end{equation}
The first few terms of this expression reproduce the expansion of $F(v)$
given in terms of $\epsilon$ earlier (Eq. \ref{fehveq}).  Similarly, an
expression for $F(y)$ can be derived by writing $v$ in terms of $\delta
y \equiv y-y_0$ and integrating the relation $\partial F(y)/\partial y=
2 v$.  It is
\begin{equation}
F(y)=-\frac{\delta}{2} +2v_0y_0 +2v_0\delta y
+\frac{4v_3}{5y_2^{3/2}}\delta y^{5/2} +{\cal O}(\delta y^{7/2})
\label{joel3}
\end{equation}
Again, this confirms the expansion given earlier (Eq. \ref{fexyeq}).

The appearance of the terms $\delta v^{5/3}$ and $\delta y^{5/2}$
means that both $F(x,y)$ and $F(h,v)$ have
non-analyticities at the first-order transition, in addition to the
expected discontinuities in their derivatives.  Thus, the traditional
mean-field
picture of a first-order phase transition as two branches of the free
energy simply crossing is not valid.  As the transition is
approached from the incompletely polarized phase, the free energy becomes
singular, and there is no obvious way to extend this branch of the free
energy into the regions $C^\pm$ in
Fig.\ \ref{fexy} to find metastable states.  It is thought
that the existence of non-analyticities at coexistence boundaries is a
generic feature;$^{(\cite{Binder})}$
the two-dimensional Ising model, {\it e.g.},
shows a weak essential singularity in $F(h)$ at $h=0$ and $T<T_c$.
Vector spin models (where, as here, a whole continuous family of order
parameter values coexists at a single point)
show power law singularities similar to the ones
found here, albeit with different values for the exponents (we are not
aware of any previous observation of, {\it e.g.}, a ``susceptibility''
$\partial^2 F(v)/\partial v^2$ with an exponent of $-1/3$).  As a
further comparison we mention that at the PT transition, where
$y$ goes to 1, the free energies behave like$^{(\cite{JayaprakashSaam})}$
\begin{eqnarray}
F(y)&=&A+B(1-y)+C(1-y)^3+\cdots\nonumber\\
F(v)&=&\alpha +\beta \delta v^{3/2}+\gamma \delta v ^2 +\cdots
\label{PTscaling}
\end{eqnarray}

Note that the conical points are reached either by letting $a \rightarrow -\pi$
or
by letting $b \rightarrow \pm\infty$ (or some combination of both), as was
remarked earlier.  Taking $a \rightarrow
-\pi$ brings us into the point along some nonzero angle with respect to the
PT line in $(h,v)$ space, whereas $b \rightarrow \pm\infty$
takes us in tangent to the PT line.
The scaling properties derived above [Eqs. (\ref{joel1}--\ref{joel3})]
correspond to the
former case. In the special case of a trajectory tangent to the PT line,
the scaling would be determined by the limit $b \rightarrow \pm\infty$.  In
particular,
we know that the free energy $F(h,v)$ is analytic in $h$ and $v$ along
the PT line itself ($b\rightarrow\infty$ with $a=0$).  By coming into the
conical
point along other curves which are tangent to the PT line, intermediate
scaling forms of various kinds could be obtained, but such scaling behavior is
not generic and therefore somewhat contrived.

This brings us to a closely--related point which is: In which region is
the
behavior near the conical point dominated by the proximity to the conical
point, and in which region is it dominated by the proximity to the PT
lines?  For example, we might ask how to determine the region over
which we can observe the scaling ({\it e.g.} the Pokrovsky--Talapov behavior)
associated with the boundary of the completely polarized region.
The preceding discussion suggests that this region
is very narrow.  Indeed, by considering when the second term in the denominator
of (A19) of ref.\ \cite{JayaprakashSaam} is of order 1, we find that a
curve of the form $d_f \sim d_c^2$ separates the two regions.  Here,
$d_c$ is the distance in the $h$--$v$ plane to the conical point and $d_f$
is the perpendicular distance to the PT line.  Note that this curve
is tangent to the PT line, deviating only quadratically as one moves
away from the conical point.  Application of (A19) in ref.\
\cite{JayaprakashSaam} also tells us that the coefficients of the $(1-y)$ and
$(1-y)^3$ terms of the expansion of the free energy about the
completely polarized region [as in Eq.\ (\ref{PTscaling})] are
unchanged by the presence of the conical point, and thus that the coefficient
of the PT scaling term is not modified.  However, higher--order terms
in the expansion diverge as one approaches the conical point, representing
the fact the the PT scaling will be observable over an ever--smaller angular
region near the PT line.

We can also examine the temperature dependence of various quantities.
For example, the jump in $y$ along $h=-v$ in the conical point, $\Delta
y\equiv y_0(b=(\phi_0+\nu)/2)= \tanh((\phi_0-\nu)/4)$ behaves as
follows:  For $T\rightarrow 0$, {\it i.e.}\ $\Delta\rightarrow \infty$, we
find that $\phi_0-\nu\sim\beta(\delta-2\varepsilon)\rightarrow\infty$,
so that $\Delta y\approx 1 -2e^{-\beta(\delta-2\varepsilon)/2}$.  For
$T\rightarrow T_c$, {\it i.e.} $\Delta \rightarrow 1$, we find $\phi_0-\nu
\propto \nu$, and $\nu\propto\Delta^{1/2}\propto (T_c-T)^{1/2}$, so that
$\Delta y\propto (T_c-T)^{1/2}$.  Qualitatively
the same temperature dependences are
found for the opening angle $2\theta_0$
between the two boundaries of the incompletely polarized
phase at the conical point; $\theta_0$ can be obtained from
Eq. (\ref{angle}) by letting $b\rightarrow\infty$, which gives
$\tan\theta_0= \tanh((\phi_0-\nu)/2)$.

Finally,
we might ask whether there are any analytic approximations we can make
which are valid over the entire phase diagram.  Indeed, we can perform an
expansion of Eqs.\ (\ref{Req}--\ref{yheq}) to lowest nontrivial order in the
temperature.$^{(\cite{Henk})}$  To do this, we ignore terms in $\xi(u)$,
$K(u-v)$, $p^0(u)$, and $\Phi^{L,R}(u)$ which are down by factors of
order $e^{-2\nu}$ (or higher powers) from the dominant terms.
Since to this order the kernel $K(u-v)$ is identically 1,
the integral equation can then be solved analytically and
we can write down explicit expressions for $h$, $v$, $x$, and $y$ in terms
of $a$ and $b$.  The expressions for $h$ and $v$, however, involve an
integral which apparently must be evaluated numerically.  Furthermore,
even the expressions for $x$ and $y$ cannot be inverted to eliminate the
parameters $a$ and $b$ from the equations entirely (leaving us, as we had
hoped, with direct relations
between $h$, $v$, $x$, and $y$).  Therefore, these expressions, while
providing us with a simpler numerical task than solving integral equations,
do not lead to any profound simplification.

Since we have ignored terms down by orders of $e^{-2\nu}$, the
low--temperature approximation is quite accurate even up to a large fraction of
the critical temperature. One particularly interesting fact about the
accuracy of this approximation should be noted:
Although the general features, such as the locations of the PT lines
(corresponding to $a \rightarrow 0$) are not obtained exactly, everything
involving the
conical points ({\it i.e.,} the limit $a \rightarrow -\pi$ and the limit
$b\rightarrow\pm\infty$)
{\it is} exactly correct.  This includes the location of these points in the
$(h,v)$ plane, the expressions for $l_\pm$, and even the expansions for $x$,
$y$, $h$, and $v$ about the conical point [Eqs.\
(\ref{coefexp}--\ref{coef1})] to the orders we
know them.  (In particular, this means that the
transition temperature $T_c$ is obtained exactly---a curious fact for the
first term of a low--temperature expansion!)  We hypothesize that the
fact that the behavior at the conical points is obtained exactly
could be evidence that this behavior is controlled by a
zero--temperature fixed point.

\section{Analytic results for $\Delta =1$}
\label{delta1}

The case $\Delta= 1$ corresponds to the critical point $T=T_c$, where
the first-order line in Fig.~\ref{fehv} has shrunk to zero length (see
Fig.~\ref{somethign}), and the two conical points at the end of this
line have now coalesced into one point, which will disappear for $\Delta
< 1$.
 Here, it is possible to obtain some analytic results
for $a\rightarrow -\infty$, which corresponds to this limiting
point.  It turns out to be much harder to make an
expansion about $a=-\infty$ than about $a=-\pi$, and we were only able
to obtain a limited amount of information about this case.

We can go through the same steps as in section~\ref{deltalarger},
using the relevant
expressions from table~\ref{ftable}.  Now equation (\ref{Req}) can be
solved by means of a Fourier transform, defined as
\begin{equation}
\widehat{R}_0(t)=\frac{1}{2\pi}\int_{-\infty}^{\infty}e^{iut} R_0(u)du
\end{equation}
The Fourier transform of $K$ is again
the same as in ref.\ \cite{Nolden}, and the one
for $\xi$ can easily be calculated:
\begin{eqnarray}
\widehat{K}(t)&=& e^{-|t|}\nonumber\\
\widehat{\xi}(t)&=&\left\{
\begin{array}{ll}
\left\{ \begin{array}{ll}
0 & (t \geq 0)\\
2e^{bt}\sinh(t/2) & (t<0)
\end{array} \right. & (b > 1/2)\\
\left\{ \begin{array}{ll}
-2e^{bt}\sinh(t/2) & (t \geq 0)\\
0 & (t<0)
\end{array} \right. & (b< -1/2)
\end{array} \right.
\end{eqnarray}
Solving (\ref{Req}) for $\widehat{R}_0(t)$ gives
\begin{eqnarray}
\widehat{R}_0(t)&=&\left\{
\begin{array}{ll}
\left\{ \begin{array}{ll}
0 & (t > 0)\\
-e^{(b-1/2)t}& (t<0)
\end{array} \right. & (b > 1/2)\\
\left\{ \begin{array}{ll}
-e^{(b+1/2)t}& (t > 0)\\
0 & (t<0)
\end{array} \right. & (b< -1/2)
\end{array} \right.
\end{eqnarray}
As usual, $\widehat{R}_0(0)$ is not determined.  However, that does not
prevent us from Fourier transforming back to find $R_0(u)$,
\begin{equation}
R_0(u)=\left\{
\begin{array}{ll}
-\int_{-\infty}^0e^{-iut+t(b-\frac{1}{2})}dt =\frac{-1}{b-
\frac{1}{2} -iu}&~~~~(b>\frac{1}{2})\\
-\int_{0}^{\infty}e^{-iut+t(b+\frac{1}{2})}dt =\frac{1}{b+
\frac{1}{2} -iu}&~~~~(b<-\frac{1}{2})\\
\end{array}\right.
\end{equation}

It is, to a certain extent, possible to expand $R(u)$ in $1/a$ around
$a=-\infty$.  It can be shown that the leading corrections are of the
form
\begin{equation}
R(u)=R_0(u)+\frac{f(u/a)}{a^2}+i\frac{g(u/a)}{a}+ \cdots
\label{d1exp}
\end{equation}
with $f$ and $g$ real functions, $f$ even and $g$ odd.  These two
functions are determined by two integral equations which still contain
$a$.  Even though numerical evidence shows that $f$ and $g$ have
well-defined limits for $a\rightarrow-\infty$, there is no obvious way to
take this limit in the integral equations, and we have not been
able to obtain closed-form expressions for the two functions.

To get $y_0$ and $h_0$ we again apply (\ref{yheq}).
It turns out that we have to be slightly careful in taking the limit
$a\rightarrow-\infty$, because we have to consider the possibility that
$b$ diverges along with $a$, so we cannot just set $a=-\infty$ as we did
before.
{}From table~\ref{ftable} it is apparent that $p^0(a)\rightarrow0$
no matter how $a$ and $b$ go to infinity.  For the integral term we find
\begin{eqnarray}
\frac{1}{2\pi}\int_{a}^{-a}\Theta(a-v)R_0(v)dv&&\rightarrow
\frac{1}{2\pi}\Theta(-\infty)\int_a^{-a} R_0(v)dv
\nonumber\\
&&= \arctan(\frac{a}{1\mp 2b})
\end{eqnarray}
with the upper (lower) signs for $b>1/2$ ($b<-1/2$).
So we find that for $a\rightarrow-\infty$
\begin{equation}
y_0=\left\{
\begin{array}{ll}
1-\frac{2}{\pi}\arctan(-a/2b) & (b>1/2)\\
1-\frac{2}{\pi}\arctan(a/2b) & (b<-1/2)
\end{array}
\right.
\label{d1y0}
\end{equation}
and $h_0=0$ in both cases.  So by letting $b$ go to infinity along with
$a$, any positive value of $y_0$ can be selected.
It follows from the expansion (\ref{d1exp}) that the corrections to both
$h$ and $y$ will be of order $1/a$; however, the coefficients of these
terms cannot be
calculated without knowledge of the functions $f$ and $g$.

To find expressions for $x$ and $v$ we can again use the symmetries
discussed in section \ref{symms}, if every occurrence of $\nu$ is
replaced by $1/2$.  In particular, we immediately find that $v_0=0$.

It is also possible to find the free energy in the limit that
$a\rightarrow-\infty$.  As before, it turns out that for $b<-1/2$
and $b>\phi_0$ the right eigenvalue is largest, and for $1/2 <
b<\phi_0$ the left eigenvalue is largest.  In all cases we find that
$F_0(h,y) = -\delta/2$, independent of $b$.  This
confirms the above conclusion that $v_0=\partial F_0/\partial y_0 = 0$.

We can again examine the scaling behavior of the
free energies.  As in the
case $\Delta >1$, $F(h,y)$ is a regular power series in
the expansion variable, $1/a$.  However, in
contrast to that case, the expansions
for $y$ and $h$ both contain linear terms in the expansion variable.
This implies that the polarizations are non-singular functions of the
fields.
Doing the Legendre transform
of $F(h,y)$ to obtain $F(x,y)$, we find that the
coexistence regions $C^{\pm}$ have shrunk to the line $x=y$, and the behavior
around this line is now quadratic in $x-y$.  So both the groove at $x=y$
and the singularities at the lines $\ell_{\pm}$ (which are now
collapsed to $x=y$) have disappeared.  The free energy can thus be
written
\begin{equation}
F(x,y)=-\frac{\delta}{2} +A(x-y)^2 + {\cal O}((x-y)^3)
\end{equation}
where the coefficient $A=A(x+y)$ is a function of the position along
$x=y$.  Similarly, the free energy as a function of the fields is given
by
\begin{equation}
F(h,v)=-\frac{\delta}{2} - y_0 (h+v) +{\cal O}((v-h)^2,(v+h)^2)
\end{equation}
where $y_0$ depends on the ratio $a/b$ according to Eq.\ (\ref{d1y0}),
{\it i.e.}, it depends on the details of how the point $h=v=0$ is
approached.  Note that both the jump in the slope of $F(h,v)$ on
entering this point, and the singularities on approaching it, have
disappeared; however, the slope in the $h+v$ direction
(given by $-y_0$) still depends on the way in which this point is
approached.  This reflects the fact that,
as $T \rightarrow T_c$ from above, the two PT boundaries
approach each other, and the curvature of $F(h,v)$ in the $h+v$
direction goes to infinity.

\section{Conclusion}
\label{concl}

To summarize, we have studied the phase diagram of the asymmetric,
ferroelectric six-vertex model in the low-temperature phase.  From
numerical results we get a global picture of the behavior of the free
energy as a function of the polarizations and fields, $F(x,y)$ and
$F(h,v)$, and by analytical methods we found several new and interesting
features.  For $T<T_c$ ($\Delta > 1$), in addition to the groove that
was already known to exist in the free energy surface $F(x,y)$, there
are additional singularities along the lines $\ell_{\pm}$, corresponding
to two conical points in $F(h,v)$.  In these points,
which lie at the ends of a ridge in $F(h,v)$, all polarizations
on $\ell_{\pm}$ are stable for one value of the fields $h$ and $v$. Also
associated with these points are the two coexistence
regions $C^{\pm}$, where
mixtures of the coexisting states are stable.

As $T\rightarrow T_c$ ($\Delta \rightarrow 1$), the size of the regions
$C^{\pm}$ decreases, as does the length of the ridge in $F(h,v)$.
At $T=T_c$, the
two conical points merge as the ridge is reduced to a point, and the
regions $C^{\pm}$ vanish.  There is still a family of polarizations,
$x=y$, that is stable in the (one) conical point.  If there are any
non-analyticities left in the free energy they are of higher order than
the terms we have been able to study.  Above $T_c$, the
conical point disappears completely.

The behavior of the free energies when approaching the conical points is
rather unexpected.  Both the fact that there {\em are} power law
divergences,
while the phase transition is discontinuous and thus first-order, and
the specific exponents of the power laws, are interesting.
It would be useful to find
metastable states in the regions $C^{\pm}$ with a uniform polarization,
and examine how such a state decays into a mixture of stable states on the
lines $\ell_{\pm}$.  However, because of the singularities in $F$ there
is no obvious way to analytically continue the free energy into the
unstable region, in order to find such metastable states in a
straightforward manner.
Also, there does not seem to be any mechanism to
select one of the many possible mixtures that such a state might break
up into.  Thus one can ask how the process of phase separation, or
spinodal decomposition, would actually take place.

A particular system in which this issue needs to be addressed is the
phase separation of crystal surfaces.$^{(\cite{PhaseSeparation,JDDJ})}$
Through the mapping of the
six-vertex model onto restricted solid-on-solid
models$^{(\cite{vBeijeren,JayaprakashSaam})}$
the polarizations $x$ and $y$ are found
to be equivalent to surface orientations, and the free energy $F(h,v)$
turns out to be a replica of the crystal shape
itself,$^{(\cite{JayaprakashSaam,Andreev})}$
while $F(x,y)$ is the surface tension.
Thus, the conical points represent actual points on the crystal surface
that have a conical geometry.\footnote{For $\Delta <-1$, the free energy
as a function of $x$ and $y$ has a conical point
({\it i.e.}\ there is a cusp in the surface tension); this leads to a flat
region in the crystal shape $F(h,v)$: a facet.$^{(\cite{Nolden,NoldenT})}$
This
situation is thus the converse of what takes place for $\Delta >1$.}
In these points, various surface orientations all coexist.  Also, there
is a discontinuity in orientation as one crosses a conical point, since
the orientations that correspond to states in the regions $C^{\pm}$ are
not stable anywhere.  Furthermore, if a crystal is cleaved
to form a surface having one of these unstable orientations
(say in $C^+$), it
should decompose into a mixture of stable
orientations selected from the set $\ell_{+}$,
which then coarsens over time.  This is the phase
separation process mentioned in the previous paragraph.  These issues
are discussed in more detail in ref.\ \cite{JDDJ}.

\section*{Acknowledgements}

We are grateful to Henk van Beijeren, Mark Holzer,
Marcel den Nijs, Ole Warnaar, and Michael Wortis for useful discussions.
This work was supported by the NSERC of Canada.

\appendixon

\namedappendix{The Fourier coefficients of $\xi(u)$}
\label{xi}

The Fourier coefficients of $\xi$ are given by
\begin{equation}
\widehat{\xi}_n = \frac{1}{2\pi} \int_{-\pi}^{\pi} du \xi(u) e^{inu}
= \frac{1}{2\pi} \int_{-\pi}^{\pi} du \frac{e^{inu}\sinh\nu}{
\cosh \nu -\cos(u+ib)}
\end{equation}
For $n\geq 0$, substituting $z=e^{iu}$ gives
\begin{equation}
\widehat{\xi}_n = -i\frac{\sinh\nu}{\pi}\int_{\bigcirc} dz \frac{z^n}{
2z\cosh \nu -e^b-z^2e^{-b}}
\end{equation}
The contour is a counterclockwise circle with radius 1.
The poles are at $z_1=e^{-\nu+b}$ and $z_2=e^{\nu+b}$.
For $b>\nu$ both poles lie outside the contour,
and $\widehat{\xi}_n=0$.  For $b<-\nu$ they are both inside
the contour, and the integral is
\begin{equation}
2\pi i \frac{-i\sinh \nu}{\pi} \left[ \frac{z_1^n}{2\sinh \nu} + \frac{
z_2^n}{-2\sinh \nu}\right]
= -2e^{bn} \sinh n\nu
\end{equation}
So for $n\geq 0$
\begin{equation}
\widehat{\xi}_n=\left\{
\begin{array}{ll}
0 & (b>\nu)\\
-2e^{bn}\sinh(n\nu) & (b<-\nu)
\end{array} \right.
\end{equation}
For $n<0$, we substitute $z=e^{-iu}$ instead, and in a similar way we
find
\begin{equation}
\widehat{\xi}_n=\left\{
\begin{array}{ll}
2e^{bn}\sinh(n\nu) & (b>\nu)\\
0 & (b<-\nu)
\end{array} \right.
\end{equation}

\namedappendix{The integrals $I_n$}
\label{eye}

The integrals $I_n$ are defined as
\begin{equation}
I_n=-\int_{-\pi}^{\pi} \Theta(-\pi-v)e^{-inv} dv
\label{integrali}
\end{equation}
The function $\Theta(-\pi-v)$ can be rewritten as
\begin{equation}
\Theta(-\pi-v)
= i\left\{ \ln\frac{-e^{iv}-e^{-2\nu}}{e^{iv}+e^{2\nu}}+2\nu
\right\}
\label{theta}
\end{equation}
where the $\ln$ has a branch cut along $[0,\infty)$, so that its
imaginary part runs from $0$ to $2\pi i$.  This is because
$\Theta(-\pi-v)$ must run from $0$ for $v=-\pi$ to $-2\pi$ for
$v=\pi$.

For $n\leq 0$ we use (\ref{theta}), define $z=-e^{iv}$, so that
\begin{equation}
I_n=-(-1)^n \int_{\bigcirc} dz \left\{\ln\frac{z-A}{B-z} +C\right\}
z^{|n|-1}
\end{equation}
with $A=e^{-2\nu}$, $B=e^{2\nu}$, $C=2\nu$.
The contour is given in Fig.~\ref{contour1}.
So we must examine integrals of the type
\begin{equation}
\widetilde{I}_n = \int_{\bigcirc} \left\{\ln\frac{z-A}{B-z} +C\right\}
z^{n-1}dz
\end{equation}
for $n\geq 0$.
For $n \neq 0$
\begin{eqnarray}
\widetilde{I}_n &+& \int_A^1\left[\ln\left|\frac{x-A}{B-x}\right| +C +
\gamma_1\right]
x^{n-1} dx \nonumber \\
&&+ \int_1^A\left[\ln\left|\frac{x-A}{B-x}\right| +C +
\gamma_2\right] x^{n-1} dx + \int_{\circ} {\rm little~loop} =0
\label{intt}
\end{eqnarray}
where now the $\ln$'s are the principal branch ({\it i.e.} with a branch cut
along $(-\infty,0]$),
the $\gamma$'s are the imaginary terms that result from approaching the
cut from above or below.  Since the integral around
the little loop goes to zero, and the line integrals give
$(1-A^n)(\gamma_1-\gamma_2) /n$, we find
\begin{equation}
\widetilde{I}_n=\frac{2\pi i}{n}(1-A^n)
\end{equation}
For $n=0$ there is an extra pole at $z=0$, so we must replace the right
hand side of (\ref{intt}) with
\begin{equation}
2\pi i(\ln\frac{-A}{B} +C)=2\pi i(\ln A
-\ln B + i\pi +C)
\end{equation}
The two line integrals give $-(\gamma_1-\gamma_2)\ln A = 2\pi i \ln A$.
So now
\begin{equation}
\widetilde{I}_0 = -2\pi^2 +2\pi i(C -\ln B)
\end{equation}

For $n<0$ a similar calculation is performed by defining $z=-e^{-iv}$.
The end result for $I_n$ is
\begin{equation}
I_n=\left\{
\begin{array}{ll}
(-1)^n \frac{2\pi i}{n}(1-e^{-2n\nu}) & (n>0)\\
2\pi^2 & (n=0)\\
(-1)^n \frac{2\pi i}{n}(1-e^{2n\nu}) & (n<0)
\end{array}\right.
\end{equation}

\namedappendix{The integrals $J_n$}
\label{jay}

Wanted are integrals of the form
\begin{equation}
J_n=\int_{\bigcirc} dz\left[ \ln\frac{z-A}{z-B} +C\right] z^{n-1}
\label{integralj}
\end{equation}
where $n\geq 0$, $A,B>0$, and the $\ln$ is the principal branch.
There are six different cases, corresponding
to the six orderings of $A$, $B$, and $1$.
They can all be computed
in a similar spirit as in~\ref{eye}.
As an example, the contour for the case $B<A<1$ is shown in
Fig.~\ref{contour2}.  As another example, the contour from
\ref{eye} can be used for the case $A<1<B$, provided that the
values $0$ and $2\pi$ for the imaginary part of the logarithm along the
branch cut are replaced by $-\pi$ and $\pi$, respectively.
The results for $J_n$ are displayed in table~\ref{itable}.

\namedappendix{The zeroth order free energy}
\label{fe}

In this appendix we calculate the free energy (\ref{feeq}),
to zeroth order in $\epsilon$, for both
the left and right eigenvalue, and compare them to find the larger
eigenvalue ({\it i.e.} lower free energy).
There are four cases:

\subsubsection*{Case 1: $b>\nu$, $\Phi=\Phi^R$}

In this case, $R_0$ is given by (see Eqs.\ (\ref{fc}) and (\ref{fred}))
\begin{equation}
R_0(u)=\widehat{R}_0 -\sum_{n=-\infty}^{-1} e^{(b-\nu)n} e^{-inu}
\end{equation}
Substituting this in Eq. (\ref{feeq}) gives
\begin{eqnarray}
-\beta F_R(h,y)&=& \frac{1}{2}(\ln \eta +2\beta h) \nonumber\\
&&-\frac{1}{2\pi} \sum_{n=-\infty}^0 \int_{-\pi}^{\pi} \Phi^R(u)
e^{-inu} du \widehat{R}_n
\label{fer}
\end{eqnarray}
Now, defining $z=e^{iu}$, and using~\ref{jay},
\begin{eqnarray}
\int_{-\pi}^{\pi}\&\& \Phi^R(u) e^{-inu}du =
 -i\int_{\bigcirc} \left\{ \ln\frac{z -e^{2\nu - \phi_0 +b}}{z -e^{-\phi_0
+b}} -\nu \right\}z^{-n-1} dz\nonumber\\
\&\&= \left\{
\begin{array}{ll}
0& (b>\phi_0)\\
-\frac{2\pi}{n}(e^{-n(b-\phi_0)}-1) & (-2\nu + \phi_0 < b< \phi_0)\\
-\frac{2\pi}{n} ( e^{-n(b-\phi_0)}-e^{-n(2\nu+b-\phi_0)}) & (b<
-2\nu+\phi_0)
\end{array}
\right. ~~~~(n<0)\nonumber\\
\&\&= \left\{
\begin{array}{ll}
2\pi\nu & (b>\phi_0)\\
2\pi(\nu-\phi_0+b) &  (-2\nu + \phi_0 < b< \phi_0)\\
-2\pi\nu & (b<-2\nu+\phi_0)
\end{array}\right.~~~~~(n=0)
\label{ints}
\end{eqnarray}
Substituting (\ref{ints}) into (\ref{fer}) gives
\newline\newline
a)  $\nu<b<-2\nu+\phi_0$
\newline
\begin{eqnarray}
-\beta F_R(h,y)
=\frac{1}{2}(\ln\eta\&\&+2\beta h)+\nu\widehat{R}_0 -\ln(1-e^{\nu-\phi_0})
+\ln(1-e^{3\nu-\phi_0})
\end{eqnarray}
b) $-2\nu+\phi_0<b<\phi_0$
\newline
\begin{eqnarray}
-\beta F_R(h,y)
\&=\& \frac{1}{2}(\ln\eta+2\beta h)+
(\phi_0-\nu-b)\widehat{R}_0 -\ln(1-e^{\nu-\phi_0})
+\nonumber\\
&&\ln(1-e^{\nu-b})
\end{eqnarray}
c) $b>\phi_0$
\begin{equation}
-\beta F_R(h,y)
= \frac{1}{2}(\ln\eta+2\beta h)
-\nu\widehat{R}_0
\end{equation}

\subsubsection*{Case 2: $b>\nu$, $\Phi=\Phi^L$}

A very similar calculation gives
\newline\newline
a)  $\nu<b<\phi_0$
\newline
\begin{eqnarray}
-\beta F_L(h,y) &=&
\nonumber \\
&=&-\frac{1}{2}(\ln\eta+2\beta h)-\nu\widehat{R}_0 -\ln(1-e^{\nu-\phi_0})
+\ln(1-e^{-\nu-\phi_0})
\end{eqnarray}
b) $\phi_0<b<2\nu+\phi_0$
\newline
\begin{equation}
-\beta F_L(h,y)
= -\frac{1}{2}(\ln\eta+2\beta h)
-(\phi_0+\nu-b)\widehat{R}_0 -\ln(1-e^{\nu-b})
+\ln(1-e^{-\nu-\phi_0})
\end{equation}
c) $b>2\nu+\phi_0$
\begin{equation}
-\beta F_L(h,y)
= -\frac{1}{2}(\ln\eta+2\beta h)
+\nu\widehat{R}_0
\end{equation}

\subsubsection*{Case 3: $b<-\nu$, $\Phi=\Phi^R$}

Here we find that
\begin{equation}
-\beta F_R(h,y)
= \frac{1}{2}(\ln\eta+2\beta h)
+\nu\widehat{R}_0
\end{equation}

\subsubsection*{Case 4: $b<-\nu$, $\Phi=\Phi^L$}

Here,
\begin{equation}
-\beta F_L(h,y)
= -\frac{1}{2}(\ln\eta+2\beta h)
-\nu\widehat{R}_0
\end{equation}

{}~\newline\newline
Now we have to collect all information on $-\beta F_{R,L}(h,y)$
and find the maximum over $R,L$ for the various intervals.
The simplest case is $b<-\nu$:
Here,
\begin{equation}
-\beta F_R(h,y)
=\frac{1}{2}(\ln\eta +2\beta h)+\nu\widehat{R}_0=
\beta F_L(h,y)
\end{equation}
Also,
\begin{equation}
\widehat{R}_0=-\frac{e^{\nu+b}}{1+e^{\nu+b}} ~~~~\mbox{\rm and}~~~~
2\beta h =\nu
\end{equation}
We also know that $\ln \eta>\nu$, say $\ln\eta =\nu+\zeta$ with
$\zeta\geq 0$.
Then
\begin{equation}
-\beta F_R(h,y)
=\nu\frac{1}{1+e^{\nu+b}}+\frac{1}{2}\zeta>0
\end{equation}
So $ -\beta F_R(h,y) $ corresponds to the larger eigenvalue in this regime.
Then,
\begin{equation}
-\beta F_0(h,y)=\frac{\beta\delta}{2}+\nu(\widehat{R}_0 +\frac{1}{2})
=\frac{\beta\delta}{2}+\frac{\nu}{2}y
\end{equation}

For
$b>\nu$
similar reasoning shows that for $b\in[\nu,\phi_0] $ the left eigenvalue
is largest, and for $b>\phi_0$ the right eigenvalue is largest, and we
find
\begin{equation}
-\beta F_0(h,y)=\frac{\beta\delta}{2}-\nu(\widehat{R}_0 +\frac{1}{2})
=\frac{\beta\delta}{2}-\frac{\nu}{2}y
\end{equation}

\appendixoff

\newpage

\section*{Tables}

\begin{table}[h]
\begin{center}
\begin{tabular}{|l|l|l|}
\hline
{}~ & $\Delta=1$ & $\Delta > 1$\\
\hline
$\Delta =$ & 1& $\cosh \nu~~~~ (\nu>0)$\\
$e^{i p^0(u)} =$ & $\frac{1+2iu-2b}{-1+2iu-2b} $&
$\frac{e^{\nu}-e^{-iu+b}}{-e^{\nu-iu+b}+1}$\\
Range of $a$ & $(-\infty,0]$&$[-\pi,0]$\\
Range of $b$ & $(-\infty,-\frac{1}{2})\cup(\frac{1}{2},\infty)$&$
(-\infty,-\nu)\cup(\nu,\infty)$\\
$\Theta(u-v)=$& $2\arctan(u-v)$&$2\arctan\left[\coth\nu \tan\left(
\frac{u-v}{2}\right)\right]$\\
Range of $\Theta$ & $(-\pi,\pi) $& $(-2\pi,2\pi)$\\
$K(u-v)=$&$\frac{2}{1+(u-v)^2}$&$\frac{\sinh 2\nu}{\cosh 2\nu
-\cos(u-v)}$\\
$\xi(u)=$&$\frac{4}{1+4(u+ib)^2}$&$\frac{\sinh\nu}{\cosh\nu
-\cos(u+ib)}$\\
$\phi_0:$&$ \phi_0=\frac{1}{2}\frac{\eta+1}{\eta-1}$&$e^{\phi_0}=
\frac{\eta e^{\nu}-1}{\eta-e^{\nu}}$\\
$\Phi^{\rm R}(u)= $&$ \ln\frac{\phi_0-1+iu-b}{\phi_0+iu-b}$&$ \ln\left(
\frac{e^{-iu}-
e^{-2\nu+\phi_0-b}}{e^{-iu}- e^{\phi_0-b}}\right)+\nu$\\
$\Phi^{\rm
L}(u)=$&$\ln\frac{\phi_0+1+iu-b}{\phi_0+iu-b}$&$\ln\left(\frac{e^{-iu}-
e^{2\nu+\phi_0-b}}{e^{-iu}- e^{\phi_0-b}}\right)-\nu$\\
\hline
\end{tabular}
\end{center}
\caption{The definitions of various quantities and
functions for the two transformations for $\Delta \geq 1$.
Note that some of the functions depend on $b$, but the explicit
dependence has been suppressed for brevity.
We use the definitions of $\Theta$ and $K$ as they are given in ref.\
\protect \cite{Nolden}; these definitions differ from the ones used in
refs.\
\protect \cite{LiebWu,LYS,SYY} by a sign in $\Theta$ and a factor of
$2\pi$ in $K$.}
\label{ftable}
\end{table}

\newpage

\begin{table}[h]
\begin{center}
\begin{tabular}{|l|l|l|}
\hline
& $J_0$ & $J_n~~~ (n\neq 0)$\\
\hline
$1>A>B$&$2\pi i C$&$2\pi i(B^n-A^n)/n$\\
$A>1>B$&$2\pi i(\ln A+C)$&$ 2\pi i(B^n-1)$\\
$A>B>1$&$2\pi i(\ln A-\ln B+C)$&$0$\\
$1>B>A$&$2\pi i C$&$2\pi i(B^n-A^n)/n$\\
$B>1>A$&$2\pi i(-\ln B+C)$&$ 2\pi i(1-A^n)$\\
$B>A>1$&$2\pi i(\ln A-\ln B+C)$&$0$\\
\hline
\end{tabular}
\end{center}
\caption{The integrals $J_n$.}
\label{itable}
\end{table}

\newpage
\section*{Figure captions}

\begin{figure}[h]
\caption{
The six vertices and their energy assignments.
\label{vertices}}
\end{figure}

\begin{figure}[h]
\caption{
Phase diagram of the six--vertex model in the $(h,v)$ plane for the case
$\Delta > 1$. (Specific parameter values are $\delta = 1.5$,
$\epsilon = 0.3$, and $k_B T = 1.1$.)  The bold lines show the phase
boundaries.  The dashed lines show the division of the incompletely polarized
part of the phase diagram into the six regions which are discussed in the text.
\label{pdhv}}
\end{figure}

\begin{figure}[h]
\caption{
Free energy surface $F(h,v)$ and phase diagram for the six--vertex model in the
$(h,v)$ plane
for the case $\Delta > 1$. (Specific parameter values are $\delta = 1.1$,
$\epsilon = 0.45$, and $k_B T = 0.35$.)
The bold lines on both surface and base plane show the phase boundaries.
For the completely polarized phases, the values of the polarizations
$(x,y)$ are noted. All boundaries are second--order (specifically,
Pokrovsky--Talapov), except for the one between the $(x,y) = (1,1)$ and
$(-1,-1)$ phases, which is first--order.
The filled squares at each end of the first--order line are the conical
points at $(h,v)=\pm(-{\nu\over{2\beta}},{\nu\over{2\beta}})$.
\label{fehv}}
\end{figure}

\begin{figure}[h]
\caption{
A section of the phase diagram shown in the base plane of
Fig.~\protect\ref{fehv}.  The
solid lines again denote phase boundaries and the filled square denotes the
conical point.  The dashed and dotted lines are some contours of constant
$a$ and $b$, respectively.  From top to bottom, the dashed lines show
$a \approx -0.07$, $-0.20$, $-0.51$, and $-1.02$.
{}From left to right, the dotted lines
show $b = (\nu+\phi_0)/2 \approx 2.59$ (which gives $y=-x$ and $h=-v$),
$b \approx 2.72$, $b = \phi_0 \approx 2.87$ (which gives $x=0$), and
$b \approx 3.13$.
Note that the phase boundaries themselves correspond to the
limit $a \to 0$, with the value of $b$ determining the location along these
boundaries.  The conical point corresponds to either $a \to -\pi$
or $b \to \infty$.  In the former limit, the angle of approach to the conical
point is determined by the value of $b$, while in the latter limit, the
approach
is tangent to the phase boundary for all values of $a$.
\label{lines}}
\end{figure}

\begin{figure}[h]
\caption{
Free energy surface $F(x,y)$ and phase diagram for the six--vertex model in the
$(x,y)$ plane, for the same parameter values as in Fig.~\protect\ref{fehv}.
The two coexistence regions $C^\pm$ correspond to the two conical points
$(h,v)=\pm(-{\nu\over{2\beta}},{\nu\over{2\beta}})$
at each end of the first--order line in Fig.~\protect\ref{fehv}.  The regions
are bounded by the lines $l_\pm$ and $y = x$ (shown as bold lines).
The line $y=x$ is also a coexistence region, corresponding
to the entire first--order line in Fig.~\protect\ref{fehv}.
[Although it is not obvious from
this perspective, $F(x,y)$ is constant along $y=x$.]
\label{fexy}}
\end{figure}

\begin{figure}[h]
\caption{
Phase diagram of the six--vertex model in the $(h,v)$ plane for the case
$\Delta = 1$. (Specific parameter values are $\delta = 1.1$,
$\epsilon = 0.45$, and $k_B T = k_B T_c \approx 0.5913$.)
The bold lines show the phase boundaries.  For the completely polarized phases,
the values of the polarizations $(x,y)$ are noted.  The phase
boundaries are all second--order, since the first--order boundary
between the $(x,y) = (1, 1)$ and $(-1,-1)$ phases has shrunk to zero length.
The two conical points have thus coalesced into a single point at
$(h,v) = (0,0)$.
\label{somethign}}
\end{figure}

\begin{figure}[h]
\caption{Contour of integration in the complex $z$--plane
for the integrals $I_n$ of Eq.\ (\protect\ref{integrali}).
A branch cut runs along the real axis between $A$ and $B$.  The labels
$0$ and $2\pi$ above and
below this branch cut give the value of the imaginary part of the
logarithm.  For the case $n = 0$, there is also a pole at $z = 0$.
\label{contour1}}
\end{figure}

\begin{figure}[h]
\caption{Contour of integration in the complex $z$--plane
for the integrals $J_n$ of Eq.\ (\protect\ref{integralj}) for the case $B < A <
1$.
A branch cut runs along the real axis between $A$ and $B$.  The labels
$\pm \pi$ above and below this branch cut give the value of the imaginary part
of the logarithm.  For the case $n = 0$, there is also a pole
at $z = 0$.
\label{contour2}}
\end{figure}


\begin{thebibliography}{99}
\bibitem{Henk}  This idea was suggested to us by Henk van Beijeren (private
communication).
\end{thebibliography}
\end{document}